\pdfoutput=1

\documentclass[fleqn,11pt]{wlscirep}
\usepackage[utf8]{inputenc}
\usepackage[T1]{fontenc}
\title{AlphaFold Accelerates Artificial Intelligence Powered Drug Discovery: Efficient Discovery of a Novel Cyclin-dependent Kinase 20 (CDK20) Small Molecule Inhibitor}

\author[1,*]{Feng Ren}
\author[1]{Xiao Ding}
\author[1]{Min Zheng}
\author[2]{Mikhail Korzinkin}
\author[1]{Xin Cai}
\author[1]{Wei Zhu}
\author[2]{Alexey Mantsyzov}
\author[2]{Alex Aliper}
\author[2]{Vladimir Aladinskiy}
\author[1]{Zhongying Cao}
\author[1]{Shanshan Kong}
\author[2]{Xi Long}
\author[2]{Bonnie Hei Man Liu}
\author[1]{Yingtao Liu}
\author[2]{Vladimir Naumov}
\author[2]{Anastasia Shneyderman}
\author[2]{Ivan V. Ozerov}
\author[1]{Ju Wang}
\author[2]{Frank W. Pun}
\author[3]{Alán Aspuru-Guzik}
\author[4]{Michael Levitt}
\author[1,2,*]{Alex Zhavoronkov}

\affil[1]{Insilico Medicine Shanghai Ltd, Suite 902, Tower E, Changtai Plaza, 2889 Jinke Road. Pudong New District, Shanghai 201203, China}
\affil[2]{Insilico Medicine Kong Kong Ltd, 307A, Core Building 1, 1 Science Park East Avenue, Hong Kong Science Park, Pak Shek Kok, Hong Kong}
\affil[3]{Department of Chemistry, Department of Computer Science, University of Toronto, Toronto, Ontario, Canada; Vector Institute for Artificial Intelligence, Toronto, Ontario, Canada; Canadian Institute for Advanced Research, Toronto, Ontario, Canada}
\affil[4]{Department of Structural Biology, Stanford University, Palo Alto, CA, USA}

\affil[*]{Corresponding authors: Alex Zhavoronkov, email: alex@insilico.com; Feng Ren, email: feng.ren@insilico.ai}


\begin{abstract}
The AlphaFold computer program predicted protein structures for the whole human genome, which has been considered as a remarkable breakthrough both in artificial intelligence (AI) application and structure biology. Despite the varying confidence level, these predicted structures still could significantly contribute to structure-based drug design of novel targets, especially the ones with no or limited structural information. In this work, we successfully applied AlphaFold in our end-to-end AI-powered drug discovery engines constituted of a biocomputational platform PandaOmics and a generative chemistry platform Chemistry42, to identify a first-in-class hit molecule of a novel target without an experimental structure starting from target selection towards hit identification in a cost- and time-efficient manner. PandaOmics provided the targets of interest and Chemistry42 generated the molecules based on the AlphaFold predicted structure, and the selected molecules were synthesized and tested in biological assays. Through this approach, we identified a small molecule hit compound for CDK20 with a Kd value of 8.9 $\pm$ 1.6 $\mu$M ($n = 4$) within 30 days from target selection and after only synthesizing 7 compounds. Based on the available data, a second round of AI-powered compound generation was conducted and through which, a more potent hit molecule, ISM042-2-048, was discovered with a Kd value of 210.0 $\pm$ 42.4 nM ($n = 2$), within 30 days and after synthesizing 6 compounds from the discovery of the first hit ISM042-2-001. To the best of our knowledge, this is the first reported small molecule targeting CDK20 and more importantly, this work is the first demonstration of AlphaFold application in hit identification process in early drug discovery.
\end{abstract}
\begin{document}

\flushbottom
\maketitle

\thispagestyle{empty}

\section*{Introduction}

The 3D structures of proteins are highly correlated with how they function in a cell and the impacts amino acid mutations cause. A protein structure is a versatile tool to study the gene-disease association and mode of action (MoA), to evaluate the druggability as a therapeutic target. Structure-based drug discovery (SBDD) has been a mainstay method to identify hit molecules and perform lead optimization, which requires the 3D structure of a target\cite{batool_structure-based_2019,nyiri_structure-based_2020,marineau_discovery_2021}. After an endeavor of decades, only a small fraction of the known proteins have experimentally determined structures. 

Accurate protein structure prediction has been a longstanding challenge until the appearance of AlphaFold at CASP14\cite{jumper_applying_2021}. The structure folds predicted by AlphaFold achieve the accuracy level comparable with the experimental methods\cite{jumper_highly_2021,evans_protein_2021}. The scientific community celebrated  DeepMind’s accomplishment\cite{akdel_structural_2021,noauthor_ai_2021,thornton_alphafold_2021,varadi_alphafold_2022,zhang_applications_2021} and the release of proteome-wide AlphaFold DB10, which now is expanded to contain over 804K protein structures covering 21 species\cite{noauthor_alphafold_nodate}. Although AlphaFold predicted protein models have variable qualities from good, bad to ugly\cite{thornton_alphafold_2021}, the predicted local distance difference test score is provided as a confidence metrics to guide the usage of 3D structures produced by AlphaFold. The AlphaFold models have been used to aid the determination of experimental structures by crystallography\cite{flower_crystallographic_2021} and cryo-EM\cite{peter_structure_2021}, to guide the functional study of PINK1\cite{kakade_mapping_2021}, to shed light on the pathogenic mutations\cite{lin_whole_2021,sen_characterizing_2021}, and to explore the protein-protein interaction\cite{humphreys_computed_2021}. Public databases include AlphaFold models as references, e.g., UniProt\cite{noauthor_uniprot_nodate}, therapeutic target database\cite{zhou_therapeutic_2022}, and APPRIS\cite{rodriguez2022appris}. The methodology of AlphaFold has inspired RoseTTAFold\cite{baek_accurate_2021}, potentially faster and cheaper protein prediction tool with adequate accuracy and AlphaDesign\cite{jendrusch_alphadesign_2021} a protein design framework. AI-powered protein prediction has been selected as one of the 2021 breakthroughs by both Science\cite{noauthor_sciences_nodate} and Nature\cite{noauthor_science_2021}.

In this work, we rapidly identify first-in-class molecules for novel targets by combining AlphaFold predicted protein structures with the end-to-end AI-P{}owered drug discovery platforms PandaOmics and Chemistry42, starting from indication, target selection, hit generation to hit identification\cite{ivanenkov2021chemistry42}. 

While we were aware of the capabilities of the AlphaFold2 algorithm available for the scientific community, the use and modification of the algorithm for commercial purposes is still poorly understood and we used the freely-available pre-computed AlphaFold structures available freely from the AlphaFold DB repository as a starting point. 

\begin{figure}[ht]
\centering
\includegraphics[width=\linewidth]{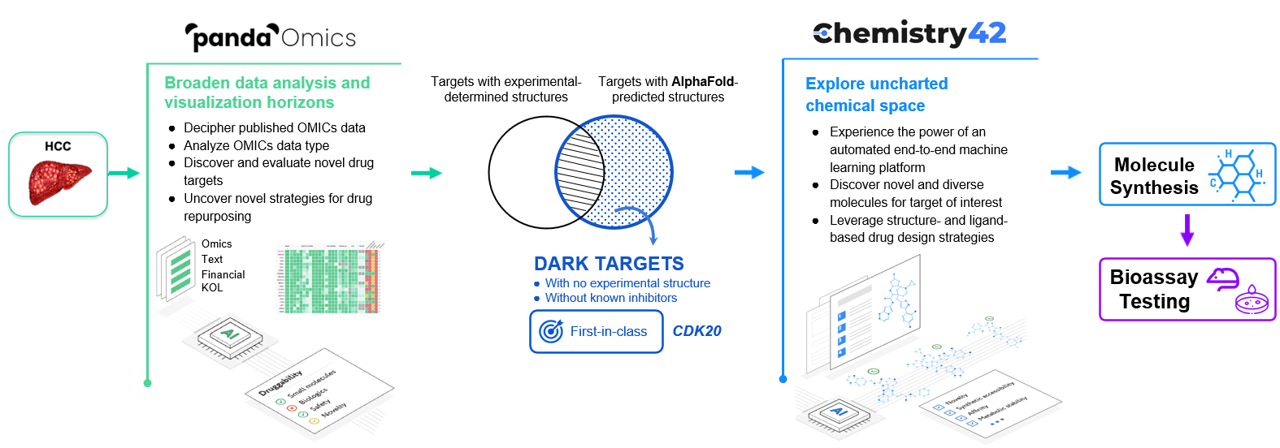}
\caption{Our approaches to combine AlphaFold with Insilico medicine end-to-end, AI-powered drug discovery platform PandaOmics and Chemistry42 in the drug discovery for hepatocellular carcinoma from target selection, hit generation to hit identification.}
\label{fig1}
\end{figure}

The general workflow was described in \textbf{Figure \ref{fig1} }and hepatocellular carcinoma (HCC) was nominated as the indication of interest due to its high prevalence in liver cancers and lack of effective treatments. In general, by analysis of text and OMICs data from 10 database for hepatocellular carcinoma, PandaOmics provides a top list of 20 targets after multiple dimensions filtration, including novelty, accessibility by biologics, safety, small molecule accessibility, and tissue specificity. CDK20 was finally selected as our initial target to work on due to its strong disease association, limited experimental structure information and with no publicly small molecule inhibitor. Through Chemistry42 structure-based drug design upon the AlphaFold predicted CDK20 structure, 8918 molecules were generated and 7 were selected for synthesis and biological testing after molecular docking, clustering, and pose inspection. Among them, compound ISM042-2-001 demonstrated a Kd value of 8.9~$\pm$~1.6~$\mu$M~($n = 4$) in CDK20 kinase binding assay and the binding mode was also docking as the guidance for further structure modifications. To the best of our knowledge, this molecule is the first reported CDK20 inhibitor and moreover, this work is also the first reported example which successfully utilized AlphaFold predicted protein structures to identify a confirmed hit for a novel target in early drug discovery.

\section*{Target Selection and Identification}

Primary liver cancer is the sixth most frequently occurring cancer and the third most common cause of cancer mortality worldwide according to GLOBOCAN 2020 update released by International Agency for Research on Cancer (IARC)\cite{batool_structure-based_2019}. Hepatocellular carcinoma (HCC) is the dominant type of liver cancer, accounting for approximately 75\% of the total patient population. The incidence rate of liver cancer is very close to its mortality rate due to very poor prognosis in all regions around the world. PD-L1 inhibitor atezolizumab in combo with bevacizumab has become the new standard-of-care (SoC) first-line treatment for advanced HCC after demonstrating a 42\% reduction in the risk of death and a 41\% reduction in the risk of disease worsening or death over the previous SoC Nexavar, but there’s still a huge unmet medical need for HCC patients.

PandaOmics is one of automated drug discovery AI engines to accelerate and optimize key steps of the early stages of drug discovery. This biocomputational platform combines bioinformatics methods for data analysis, visualization and interpretation with advanced multimodal deep learning approaches for target identification. PandaOmics therapeutic target and biomarker identification system is based on the combination of multiple scores derived from text and OMICs data associating genes with a disease of interest. Text evidence prioritization (Text, Financial and Key Opinion Leaders (KOL) score families) singles out the genes, extensively mentioned across scientific literature and grant description. OMICs-based scores, in contrast, explore the molecular connection of genes with diseases based on differential expression, gene variants, interactome topology, signaling pathway perturbation analysis algorithms\cite{ozerov_silico_2016}, knockout/overexpression experiments and more. This kind of approach allows users to unveil the hidden hypotheses that might not be obvious over common general knowledge or simple bioinformatics analysis. AI tools are extremely helpful for efficient target hypothesis generation. The overall scoring approach results in the ranked list of target hypotheses for a given disease which can be subsequently filtered according to their novelty, accessibility by small molecules and antibodies, safety, tissue specificity, crystal structure availability and major biological structures.

\begin{figure}[ht]
\centering
\includegraphics[width=\linewidth]{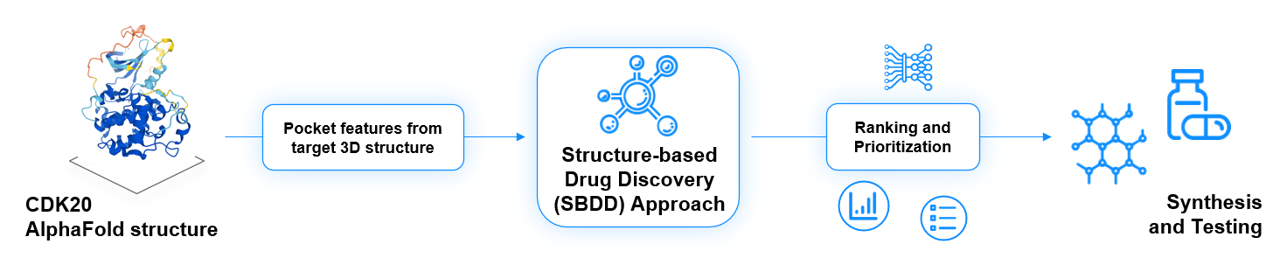}
\caption{Insilico Medicine Generative procedures for CDK20 hits.}
\label{fig2}
\end{figure}

Another unique feature of the PandaOmics platform is the ability to combine the data from different experiments into a single Meta-analysis and leverage the insights from all the datasets together for the precise target prioritization. During this study we’ve created a Meta-analysis for each of the diseases of interest composed from 10 datasets for HCC (1133 disease samples and 674 healthy controls). After obtaining the ranked list of target hypotheses we applied PandaOmics filters in order to get the list of the most promising targets that satisfy First-in-Class (see Methods section) characteristics and share the current unavailability of crystal structure but have structure folds predicted by AlphaFold.  The final list of top-20 targets was then manually curated to nominate the most promising candidates. For hepatocellular carcinoma CDK20 was chosen since it had the highest scores aligned with the First-in-class scenario. The proposed therapeutic target CDK20 was passed to Chemistry42 platform for the automated generation of small molecule inhibitors.

\section*{CDK20 as a Promising Target for Cancer Treatment }

CDK20, also known as cell cycle-related kinase (CCRK), is the latest identified member of the cyclin-dependent kinase family, which attracted great attention in recent years due to its functions (both cell cycle-dependent and -independent) in a variety of human tissues\cite{mok_ccrk_2018}. CDK20 is widely expressed at a comparable translational level in many human tissues including the brain, lung, liver, pancreas, and gastrointestinal tract\cite{uhlen_proteomics_2015}. More importantly, increasing preclinical evidence suggested that CDK20 is overexpressed in many tumor cell lines including tumor samples from patients with different types of cancer, such as colorectal cancer, hepatocellular carcinoma, lung cancer, and ovarian carcinoma\cite{an_functional_2010,feng_cell_2011,wang_cdk20_2017,wu_cell_2009}.  \textit{In vitro} studies showed that androgen receptor (AR), CDK20, and $\beta$-catenin constitute a positive feedback circuit to promote cell cycle progression in HCC cells, and CDK20 overexpression frequently correlates with ectopic expression of AR and $\beta$-catenin in primary HCC tissue samples, and with the tumor staging and poor overall survival of patients\cite{feng_cell_2011}. In lung cancer cells, CDK20 competes with nuclear factor erythroid 2-related factor 2 (NRF2) for kelch-like ECH associated protein 1 (KEAP1) binding, which prevents degradation of NRF2 and enhances its transcriptional activity, and therefore lowering the cellular reactive oxygen species (ROS) level. Moreover, CDK20 depletion in lung cancer cells demonstrates impaired cell proliferation, defective G2/M arrest, and increased radiochemosensitivity\cite{wang_cdk20_2017}. In addition to its pro-tumorigenic role through modulation of cell cycle and oncogenic signaling, CDK20 is also involved in immunosuppression in certain types of tumors. Zhou et al. reported that, by activating the EZH2-NF-$\kappa$B pathway, CDK20 expressed in HCC cells increased IL-6 production and induced immunosuppressive MDSC expansion from human peripheral blood mononuclear cells; inhibition of tumorous CDK20 increased IFN-$\gamma$\textsuperscript{+}TNF-$\alpha$\textsuperscript{+}CD8\textsuperscript{+}~T~cell infiltration and upregulated PD-L1 expression level in tumors, providing a greater chance of combination therapy with PD-L1 blockade to eradicate HCC tumors\cite{zhou_hepatoma-intrinsic_2018}.  Hence emerging scientific evidence suggested CDK20 inhibition could be considered as a promising therapeutic approach for the cancer treatment, especially for hepatocellular carcinoma (HCC).
\begin{figure}[ht]
\centering
\includegraphics[width=0.9\linewidth]{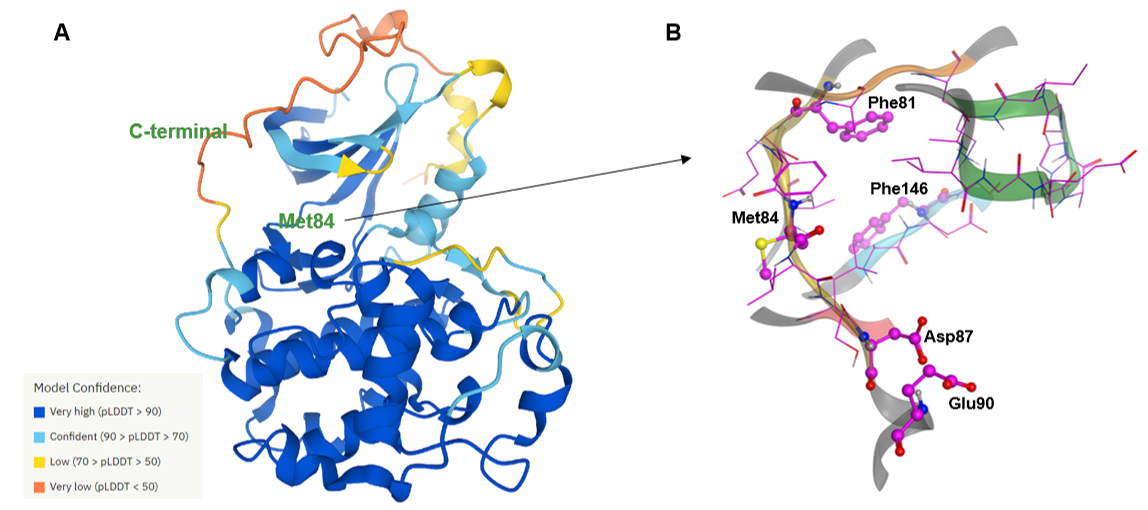}
\caption{A) The AlphaFold predicted structure of CDK20 (AF-Q8IZL9-F1-model\_v1); B) ATP pocket of CDK20 with a DFG-in (residue Phe146) conformation. Met84 is the hinge residue. P-loop is colored in green. Two acid centers Asp87 and Glu90 are located in the solvation region.}
\label{fig3}
\end{figure}

\section*{Generation of Novel Hits Targeting CDK20 by using AlphaFold Predicted Structure}
 
As of today, there is no available CDK20 inhibitor reported despite great success has been achieved for other members in CDK family. One possible reason is that there is no available 3D-structure information for this target. \textbf{Figure \ref{fig2} }described our generative procedures for the identification of CDK20 inhibitors starting from structure extracting to hit generation through SBDD approach by utilizing the generative chemistry platform Chemistry42\cite{ivanenkov2021chemistry42,zhavoronkov_deep_2019,vanhaelen_advent_2020}. 

The AlphaFold predicted CDK20 structure (AF-Q8IZL9-F1-model\_v1) has high confidence overall except for C-terminal, which could be removed since it occupies and blocks the solvation region of the ATP pocket as displayed in \textbf{Figure \ref{fig3}A}. Chemistry42 then extracted the pocket features from the target 3D structure and the system suggests CDK20 has a shallow ATP binding pocket with an estimated volume around 150 \AA\textsuperscript{3} as shown in \textbf{Figure \ref{fig3}B}. Near the hinger residue Met84, residue Phe81 occupies the gatekeeper and stops a ligand to reach the back pocket. The predicted binding pocket has a DFG-in conformation and two acid centers (Asp87 and Glu90) in the solvation region. Pocket-based generation approach was then then utilized to generate novel molecule structures. In total 8918 molecules were designed by Chemistry42. After molecular docking, clustering, and pose inspection, 54 molecules with diverse hinge core structures were prioritized and 7 compounds were selected for synthesis.

\begin{figure}[ht]
\centering
\includegraphics[width=0.8\linewidth]{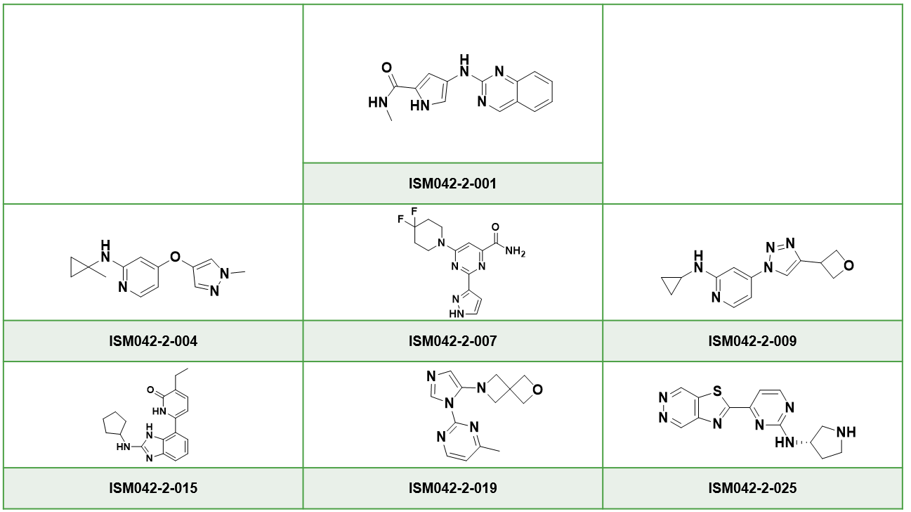}
\caption{Chemical structures for selected 7 molecules from Chemistry42 generation for synthesis and testing in CDK20 binding assay.}
\label{fig4}
\end{figure}

\section*{Results and Discussion}

\begin{figure}[ht]
\centering
\includegraphics[width=0.9\linewidth]{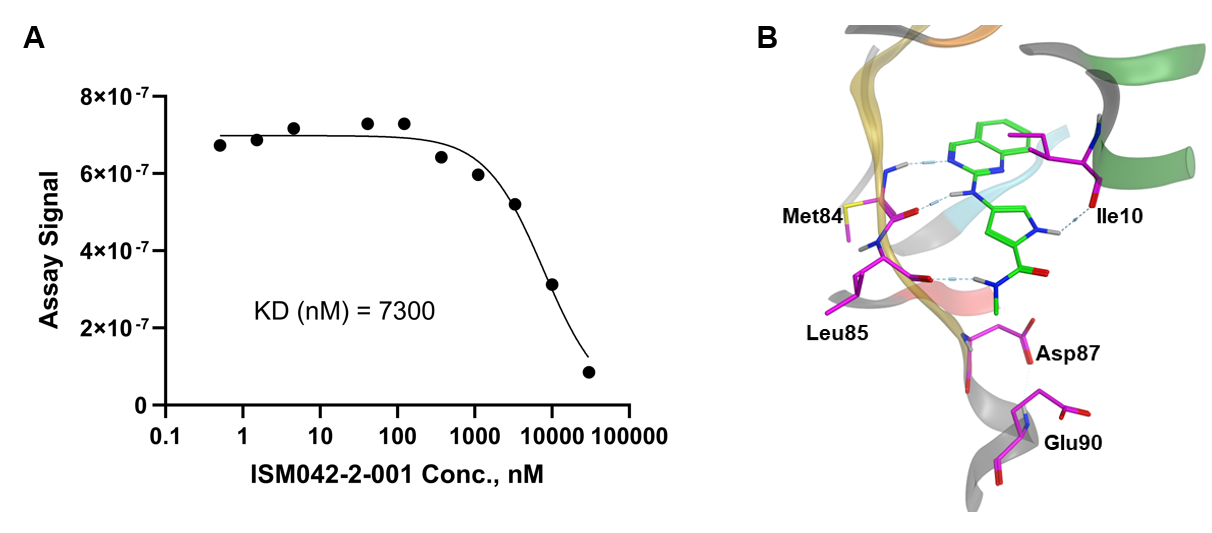}
\caption{A) One of the four binding affinity cure images for ISM042-2-001 in CDK20; B) Predicted binding pose for ISM042-2-001 with CDK20.}
\label{fig5}
\end{figure}

\textbf{Figure \ref{fig4} }shows the chemical structures for 7 compounds selected for synthesis and assessing the binding abilities towards CDK20.  The results suggested one compound ISM042-2-001 demonstrated a Kd value of 8.9 $\pm$ 1.6~$\mu$M ($n = 4$, one binding curve is shown in \textbf{Figure \ref{fig5}A}) in CDK20 kinase binding assay. We also proposed the binding mode for ISM042-2-001 as depicted in \textbf{Figure \ref{fig5}B}: four hydrogen bond interactions are represented as dash lines. Besides the two hydrogen bonds formed with the hinge residue Met84, ISM042-2-001 also interacts with residue Leu85 via the amide -NH group and with residue Ile10 in P-loop via the pyrrole -NH group. Alternatively, the amide -NH group or the pyrrole -NH group may form hydrogen bond interactions with the two acid centers Asp87 and Glu90 in the solvation region. 

\begin{figure}[ht]
\centering
\includegraphics[width=0.9\linewidth]{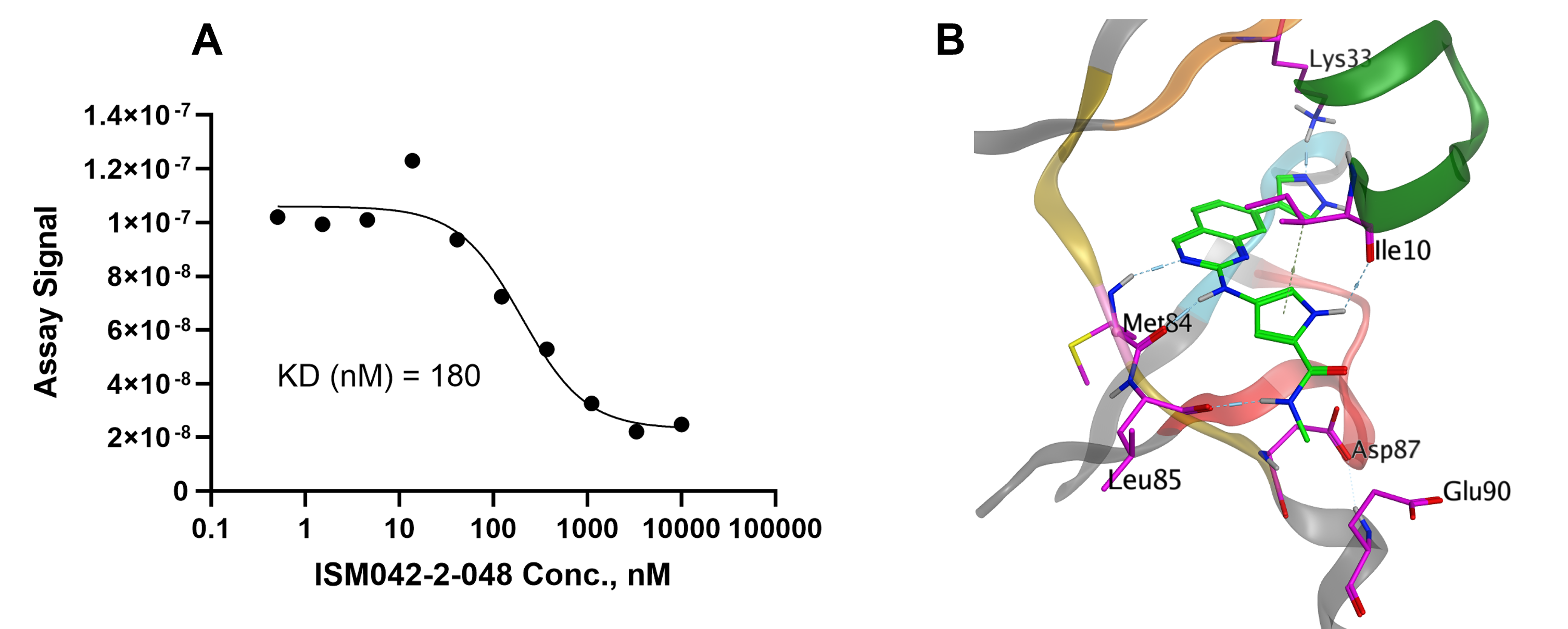}
\caption{A) One binding affinity curve for ISM042-2-048 in CDK20; B) Predicted binding pose for ISM042-2-048 in CDK20.}
\label{fig6}
\end{figure}

Based on the predicted binding pose and potency data, we conducted the second round of compound generation utilizing our generative AI tool Chemistry42. 16 novel molecules were generated aiming to improve the binding affinity based on two strategies: 1) functional groups on the quinazoline ring to occupy the hydrophobic pocket near the gatekeeper region; 2) modifications of the pyrrole-2-carboxamide group to access the solvation region and interact with acidic residues Asp87 or Glu90. 6 out of 16 generated molecules were synthesized and tested, of which ISM042-2-048 and ISM042-2-049 displayed 40 and 24 folds improvement of binding affinity comparing to ISM042-2-001, with measured Kd values of 210.0 ($\pm$ 42.4 nM) and 375 nM($\pm$ 5 nM), respectively. A predicted binding mode of ISM042-2-048 with CDK20 was showed in \textbf{Figure \ref{fig6}B}. Based on the proposed binding mode, in addition to the interactions in the hinge and solvent areas, the pyrazole group of ISM042-2-048 forms a hydrogen bonding interaction with residue Lys33, which explains the significant improvement of its binding affinity. Next-round of optimization will be initiated soon to further improve potency, and the ADME properties and kinase selectivity will also be evaluated. 

\section*{Conclusions}

Structure-based drug discovery (SBDD) has been a mainstay method to identify hit molecules and perform lead optimization and predicted protein structures by AlphaFold has been considered as a powerful tool to identify hits for novel targets with no or limited structure information. Herein, we present an example of rapid identification of a CDK20 hit molecule by combining AlphaFold with our automated drug discovery AI engines PandaOmics Chemistry42 for the treatment of HCC within 30 days covering target selection, molecule generation, compound synthesis and biological testing. Among the 7 compounds synthesized, ISM042-2-001 demonstrated a Kd value of 8.9 $\pm$ 1.6 $\mu$M ($n = 4$) in CDK20 kinase binding assay. Based on the preliminary SAR, a second round of AI-powered compound generation was conducted and 6 more compounds were synthesized and tested within 30 days from the discovery of the first hit ISM042-2-001. To our delight, a more potent hit molecule, ISM042-2-048, was discovered with a Kd value of 210.0 $\pm$ 42.4 nM ($n = 2$), which could be served as a tool molecule to evaluate biological functions for this target. Further optimization on this molecule as well as the evaluation of ADME properties and kinase selectivity are ongoing. Moreover, this work also represents an example of utilizing AlphaFold predicted protein structures for drug discovery, further studies on other target class such as GPCR and E3 ligase will be reported in due course.

\section*{Materials and Methods}

\textit{Target ID and Target proposal}: PandaOmics platform was used to conduct a hypothesis generation for Hepatocellular carcinoma, limiting the target list to the proteins whose structures were predicted by AlphaFold2. Hepatocellular carcinoma Meta-analysis combined the data from ten experiments:  GSE36376\cite{cho_expression_2020}, GSE107170\cite{cho_expression_2020,diaz_molecular_2018}, GSE102079\cite{chiyonobu_fatty_2018}, GSE45267\cite{chiyonobu_fatty_2018,wang_forfeited_2013}, GSE133039\cite{carrillo-reixach_epigenetic_2020}, GSE104766\cite{hooks_new_2018}, GSE77314\cite{hooks_new_2018,liu_potential_2016}, GSE60502\cite{wang_plasmalemmal_2014}, E-MTAB-5905\cite{losic_intratumoral_2020} and TCGA-LIHC\cite{tcga}, resulting in 1133 disease samples and 674 healthy controls.

\noindent\textit{CDK20 Human CMGC kinase binding assay}: CDK20 proteins were produced in HEK-293 cells and subsequently tagged with DNA for qPCR detection. Streptavidin-coated magnetic beads were treated with biotinylated small molecule ligands for 30 minutes at room temperature to generate affinity resins. The liganded beads were blocked with excess biotin and washed with blocking buffer (SeaBlock (Pierce), 1\% BSA, 0.05\% Tween 20, 1 mM DTT) to remove unbound ligand and to reduce non-specific binding. Binding reactions were assembled by combining kinases, liganded affinity beads, and test compounds in 1x binding buffer (20\% SeaBlock, 0.17x PBS, 0.05\% Tween 20, 6 mM DTT). Test compounds were prepared as 111X stocks in 100\% DMSO. Binding constants (Kds) were determined using an 11-point 3-fold compound dilution series with three DMSO control points. All compounds for Kd measurements are distributed by acoustic transfer (non-contact dispensing) in 100\% DMSO. The compounds were then diluted directly into the assays such that the final concentration of DMSO was 0.9\%. All reactions were performed in a polypropylene 384-well plate. Each was a final volume of 0.02 ml. The assay plates were incubated at room temperature with shaking for 1 hour and the affinity beads were washed with wash buffer (1x PBS, 0.05\% Tween 20). The beads were then re-suspended in elution buffer (1x PBS, 0.05\% Tween 20, 0.5 $\mu$M non-biotinylated affinity ligand) and incubated at room temperature with shaking for 30 minutes. The kinase concentration in the eluates was measured by qPCR. Kds were calculated with a standard dose-response curve using the Hill equation:

\noindent$\text{Response} = 
\text{Background} + \frac{\text{Signal} - \text{Background}}
{1 + (Kd^{\text{Hill Slope}} / Dose^{\text{Hill Slope}}))}$. The Hill Slope was set to -1. Curves were fitted using a non-linear least square fit with the Levenberg-Marquardt algorithm.

\noindent\textit{General procedure for the preparation of ISM042-2-001:} To a solution of ethyl 4-amino-1H-pyrrole-2-carboxylate (500 mg, 3.24 mmol) in DMF (15 mL) was added K2CO3 (672 mg, 4.86 mmol), 2-chloroquinazoline (534 mg, 3.24 mmol).  The mixture was stirred at 80 $^\circ$C for 16 hrs. The reaction mixture was washed with water (30 mL) and extracted with EtOAc (15 mL$\times$3).  The combined organic layers were washed with NaCl (30 mL$\times$2), dried over Na2SO4, filtered and concentrated. The crude was purified by column chromatography on silico gel (PE:EA~=~5:1) to give ethyl 4-(quinazolin-2-ylamino)-1H-pyrrole-2-carboxylate (632 mg, 2.24 mmol, 69.03\% yield) as a yellow solid. To a solution of ethyl 4-(quinazolin-2-ylamino)-1H-pyrrole-2-carboxylate (100 mg, 354 umol) in THF (2 mL) was added LiOH (1 M in water, 1.77 mL).  The mixture was stirred at 50 $^\circ$C for 2 hrs.  The reaction mixture was quenched by addition HCl (1 M) to neutral and then extracted with EtOAc (3 mL$\times$3). The combined organic layers were washed with NaCl (5 mL$\times$2), dried over Na2SO4, filtered and concentrated to give 4-(quinazolin-2-ylamino)-1H-pyrrole-2-carboxylic acid as yellow oil. To a solution of 4-(quinazolin-2-ylamino)-1H-pyrrole-2-carboxylic acid (100 mg, 393 umol), MeNH2 (2 M, 1.97 mL) in DMF (0.2 mL) was added CDI (76.5 mg, 472 umol) and DCC (122 mg, 590 umol).  The mixture was stirred at 25 $^\circ$C for 12 hrs. The reaction mixture was diluted with Water 2 mL and extracted with EtOAc (3 mL$\times$3).  The combined organic layers were washed with NaCl (5 mL$\times$2), dried over Na2SO4, filtered and concentrated.  The crude was purified by prep-HPLC to give ISM042-2-001 (35.0 mg, 124 umol, 31.7\% yield) as a yellow solid. LCMS: 268.0 [M+H]+. 1H NMR: (400 MHz, CDCl3): $\delta$ 9.24 - 9.02 (m, 2H), 7.85 - 7.69 (m, 3H), 7.57 - 7.47 (m, 1H), 7.43 - 7.33 (m, 1H), 6.68 (br s, 1H), 5.94 - 5.79 (m, 1H), 3.01 (d, J = 4.80 Hz, 3H). 

\section*{Acknowledgments}

Alán Aspuru-Guzik would like to thank the Canada 150 Research Chairs Program for their generous support, as well as Anders G. Frøseth.

\section*{Conflicts of interests}
Insilico Medicine is a company developing an AI-based end-to-end integrated pipeline for drug discovery and development and engaged in aging and cancer research. Alán Aspuru-Guzik is co-founder and Chief Vision officer of Kebotix, an AI-powered materials and molecular discovery company and co-founder and Chief Scientific Officer of Zapata Computing, a quantum software computing company. Alán Aspuru-Guzik is scientific advisor to Insilico Medicine.  Michael Levitt is an advisor to Insilico Medicine.  

\bibliography{main}

\end{document}